\documentclass[12pt]{article}
\usepackage{graphicx}
\usepackage{amsmath}
\usepackage{a4wide}
\newcommand{\cseveneff}{C_7^{\rm{eff}}}
\def\tp{\tilde{k}_+}
\def\tm{\tilde{k}_-}
\def\tpm{\tilde{k}_\pm}
\def\bllg{B\to\gamma\ell^+\ell^-}
\def\bwg{{B\to\gamma\ell\nu_\ell}}
\def\bgg{B\to\gamma\gamma}
\def\bqgg{B_q^0\to\gamma\gamma}
\def\dirac#1{#1\llap{/}}
\def\as{\alpha_s}
\def\eg{E_\gamma}
\newcommand{\lqcd}{\Lambda_{\textrm{\scriptsize{QCD}}}}

\begin{document}

\thispagestyle{empty}

\begin{flushright}
SHEP/02-33\\ LPT-ORSAY/02-118
\end{flushright}

\vspace{\baselineskip}

\begin{center}
\vspace{0.5\baselineskip} \textbf{\Large
Universality of Nonperturbative QCD Effects\\[0.2em]
in Radiative \boldmath{$B$}-decays}\\

\vspace{3\baselineskip}

{{\sc S.~Descotes-Genon$^{\,a,b}$} and {\sc C.T.~Sachrajda$^{\,a}$}}\\
\vspace{2\baselineskip}
\textit{$^{a}$ Department of Physics and
Astronomy, University of Southampton\\[0.1cm]
Southampton, SO17 1BJ, U.K.\\[0.2cm]
$^{b}$ Laboratoire de Physique Th\'eorique, Universit\'e Paris XI,\\[0.1cm]
B\^at. 210, 91405 Orsay Cedex, France}
\\

\vspace{1.5\baselineskip}

\textbf{Abstract}\\
\vspace{0.5\baselineskip}
\parbox{0.9\textwidth}{
We demonstrate, by an explicit one-loop calculation, that at
leading twist the nonperturbative effects in $\bwg$, $\bgg$ and
$\bllg$ radiative decays are contained in a common multiplicative
factor ($\Lambda_B(E_\gamma)$, where $E_\gamma$ is the energy of
the photon). We argue that this result holds also at higher
orders. Ratios of the amplitudes for these processes do not depend
on scales below the mass of the $B$-meson ($M_B$), and can be
calculated as perturbative series in $\alpha_s(M_B)$.}
\end{center}

\newpage

\setcounter{page}{1}


\newpage

\section{Introduction}\label{sec:intro}
In this letter we make the observation that, at leading power in
the mass of the $B$-meson ($M_B$), the nonperturbative QCD effects
in the radiative $B$-decays, $\bwg$, $\bllg$ and
$B\to\gamma\gamma$ are universal. By this we mean that, for the
same photon energy $E_\gamma$, the amplitudes for the processes
$\bwg$ and $\bllg$ are proportional to each other. For
$E_\gamma=M_B/2$ they are also proportional to the amplitude for
$B\to\gamma\gamma$. The constants of proportionality contain
kinematic factors, CKM-matrix elements and a series in
$\alpha_s(M_B)$ which is therefore calculable in perturbation
theory. Thus, in spite of the different weak-decay mechanisms for
the three processes, the strong interaction effects are common at
scales below $M_B$.

In ref.~\cite{dgs2} we showed that the two form factors
($F_{A,V}$) for the $\bwg$ decay are given, up to one-loop order,
by the \textit{factorization formula}:
\begin{equation}\label{eq:fgeneric}
F_{A,V}(E_\gamma)=\int
d\tp\Phi_+^B(\tp;\mu_F)\,T(\tp,E_\gamma;\mu_F)\,,
\end{equation}
with
\begin{equation}
T(\tp,E_\gamma;\mu_F)=
\left\{C_3^{SCET}(\mu_F)\, \frac{f_BQ_uM_B}{2\sqrt{2}E_\gamma}\right\}
\frac1{\tilde
k_+}\left[1+\frac{\alpha_s(\mu_F)C_F}{4\pi}\,K_t(\tilde
k_+,E_\gamma;\mu_F)\right], \label{eq:previous}
\end{equation}
where the momentum of the photon is in the $-$ direction,
$C_3^{SCET}(\mu_F)$ is a Wilson coefficient relating the weak
$b\to u$ current to operators of the Soft-Collinear Effective
Theory (SCET)~\cite{SCETbtou}, $f_B$ is the $B$-meson's leptonic
decay constant and $Q_u=-2/3$ is the charge of the $\bar
u$-quark~\footnote{Although we have chosen to use the SCET
formalism to sum the large logarithms, this could equivalently be
achieved using earlier approaches, in particular the
``Wilson-Line" formalism of ref.~\cite{KoSt}.}. $\mu_F$ is a
factorization scale and is conveniently chosen to be
$O(\sqrt{M_B\lqcd})$. The $B$-meson's light-cone distribution
amplitude, $\Phi_+^B$, is the component of the non-local matrix
element~\cite{GN,BF}
\begin{equation}\label{eq:phihdef}
\Phi^H_{\alpha\beta}(\tp)=\int dz_- e^{i\tp z_-}
  \left.\langle 0| \bar{u}_\beta(z) [z,0] b_\alpha(0) | H
  \rangle\right|_{z_+,z_\perp=0}\,,
\end{equation}
with the hadronic initial state $H=B$, which contributes to the
amplitudes at leading twist. In eq.~(\ref{eq:phihdef}),
$\alpha,\beta$ are spinor labels and $[z,0]$ denotes a
path-ordered exponential. Finally,
\begin{equation}\label{eq:ktdef}
K_t(\tilde k_+,E_\gamma;\mu_F)=\log^2\frac{2\sqrt{2}E_\gamma\tilde
k_+}{\mu_F^2}-\frac{\pi^2}6-1\,.
\end{equation}
The above formulae hold for photon energies satisfying
$E_\gamma\gg \lqcd$.

The nonperturbative contribution to the form factors is contained
in $\Phi_+^B(\tp)$ which is convoluted with $T$ as in
eq.~(\ref{eq:fgeneric}). The observation we make here is that
\textit{precisely} the same convolution over $\tp$ appears also in
the amplitudes for $\bllg$ and $\bgg$ decays. The
$\tp$-independent prefactor in curly brackets in
eq.~(\ref{eq:previous}) is process-dependent but perturbative.
Moreover, although the coefficient $C_3^{SCET}(\mu_F)$ in
eq.~(\ref{eq:previous}) is specific to the decay $\bwg$, the
behaviour with $\mu_F$ (for $\mu_F<m_B$) is common to all
coefficients $C_i^{SCET}$ appearing in the three radiative decays.
Thus ratios of amplitudes for different processes but with the
same $E_\gamma$ only depend on the $C_i^{SCET}$ at $\mu_F=M_B$,
and are hence calculable as perturbative series in
$\alpha_s(M_B)$. The formulae for the decay rates are presented
explicitly in sec.\ref{concs}.

Additional motivation for this study is provided by the need to
understand nonperturbative QCD effects in two-body exclusive
nonleptonic decays, for which there is a wealth of experimental
data, particularly from the $B$-factories. The demonstration of
the factorization of long-distance effects at leading twist for
these processes~\cite{BBNS1,BBNS2,BBNS3} provides the theoretical
framework for detailed phenomenological analyses. However, our
ignorance of $\Phi_+^B$ (and the lack of a similar framework for
higher-twist contributions) limits the precision with which
information about the CKM matrix elements and CP-violation can be
determined from the experimental measurements of branching ratios
and asymmetries. It is therefore important to understand whether
there are relations, similar to those presented here, also between
different nonleptonic decay amplitudes (or between contributions
to the amplitudes) which reduce or eliminate the need for a
detailed knowledge of $\Phi_+^B$. Finally we repeat that measuring
the photon's energy distribution in $\bwg$ decays may in future be
the best way of obtaining information about
$\Phi_+^B$~\cite{dgs2}.

In the next section we sketch the evaluation of the amplitudes for
$\bgg$ and $\bllg$ decays at one-loop order and the resummation of
the large logarithms. We focus on the differences with the
corresponding calculation of the $\bwg$ decay amplitude, which was
described in detail in ref.~\cite{dgs2}. The reader who is only
interested in the implications of our results can turn immediately
to section~\ref{concs} where we present the expressions for the
decay rates and discuss their significance.

\begin{figure}[t]
\begin{center}
\includegraphics[bb=80 690 190 740]{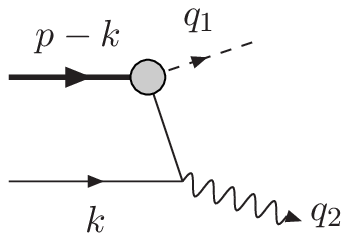}
\end{center}
\caption{Lowest-order diagram at leading twist contributing to the
process $b\bar u\to \gamma X$, where $X=\ell\nu_\ell$,
$\ell^+\ell^-$ or $\gamma$ and is represented by the dashed line.
The thick line represents the $b$-quark and the thin line a light
quark. The grey circle represents the operator responsible for the
$b\to u$ transition.\label{fig:tree}}
\end{figure}

\section{Amplitudes up to One-Loop Order}\label{sec:oneloop}

In ref.~\cite{dgs2} we have described the evaluation of the
one-loop contribution to the hard-scattering kernel for the decay
$\bwg$ in detail. The corresponding calculation for the other
radiative decays is very similar, so here we only sketch the main
points and present the results.

\begin{figure}[t]
\begin{center}
\mbox{\includegraphics[bb=80 670 190 760]{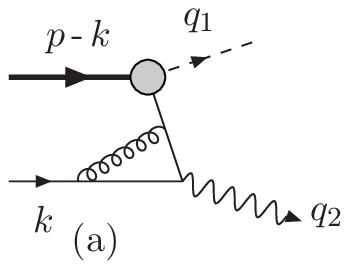}
\includegraphics[bb=80 670 190 760]{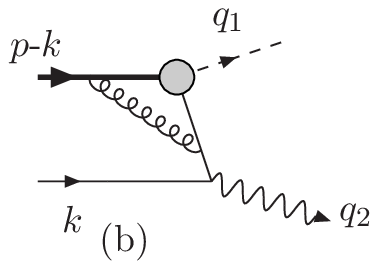}
\includegraphics[bb=80 670 190 760]{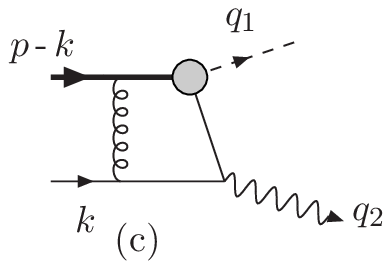}}\\
\mbox{\includegraphics[bb=80 670 190 760]{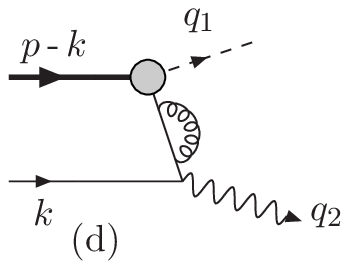}
\includegraphics[bb=80 670 190 760]{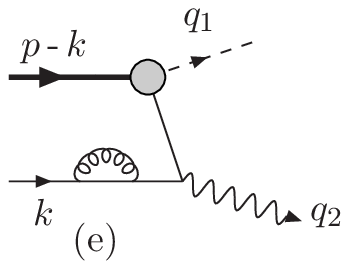}
\includegraphics[bb=80 670 190 760]{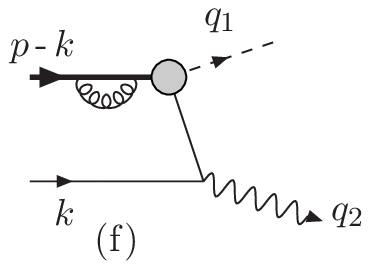}}
\end{center}
\caption{One-loop leading-twist diagrams for the process $b\bar
u\to X\gamma$.\label{fig:oneloop}}
\end{figure}

We wish to write the form factors for each of the three radiative
processes in the generic factorized form given in
eq.~(\ref{eq:fgeneric}), where the hard-scattering kernel $T$
depends on the process, but, as its name suggests, does not depend
on scales below $\mu_F$ and can be calculated in perturbation
theory. In order to determine $T$ we are free to choose any
appropriate and convenient external state, and we take a
$u$-antiquark with momentum $k$ and a $b$-quark with momentum
$p$\,--\,$k$. The leading-twist contributions are those in which
the photon is emitted from the light quark, and we will see that
the diagrams which need to be evaluated are those in
figs.~\ref{fig:tree} and \ref{fig:oneloop}.

The evaluation of the one-loop graphs in
fig.~\ref{fig:oneloop}(a),\,(d),\,(e) and (f) is common to all the
processes and is described in ref.~\cite{dgs2}. In
ref.~\cite{dgs2} we also show that the leading-twist contribution
from the box diagram in fig.~\ref{fig:oneloop}(c) comes from the
soft region of phase-space and is absorbed into the one-loop
component of the wave-function, giving no contribution to $T$.
This is also true for $\bgg$ and $\bllg$ decays. Thus in the
following we present the results for the diagrams in
fig.~\ref{fig:tree} and \ref{fig:oneloop}(b) for $\bgg$ and
$\bllg$ decays and combine them with those in ref.~\cite{dgs2} for
the other graphs (the graphs are evaluated in the Feynman gauge).

\subsection{The decay \boldmath{$\bgg$}}

The effective Hamiltonian for $\bqgg$ decays is~\cite{effham,bgg}
\begin{equation}\label{eq:heffgg}
{\mathcal{H}}= i\lambda_t^{(q)} \frac{eG_F}{4\pi^2\sqrt{2}} m_b
\cseveneff (\mu_R)\,
  [\bar{q}\gamma^\mu\gamma^\nu b_R]
  (\partial_\mu A_\nu - \partial_\nu A_\mu)\,,
\end{equation}
where $b_R=1/2(1+\gamma_5)b$, $q=d$ or $s$ and
$\lambda_t^{(q)}=V^*_{tq} V_{tb}$. $\cseveneff $ is the Wilson
coefficient at the scale $\mu_R$ (it will generally be convenient
to take $\mu_R=M_B$). The superscript {\small eff} denotes the
fact that $\cseveneff$ contains terms proportional to the Wilson
coefficients $C_{1\textrm{\,--\,}6}$ (corresponding to $\Delta
B=1$ four-quark operators) and $C_8$ (corresponding to the
chromomagnetic operator). The contributions included in
$\cseveneff$ are short-distance loop corrections to
$O_{1\,\textrm{--}\,6}$ and $O_8$ which result in an effective
vertex proportional to the operator $O_7$ at tree-level (i.e. to
the operator in eq.~(\ref{eq:heffgg})).

\begin{figure}[t]
\begin{center}
\mbox{\includegraphics[bb=80 670 190 760]{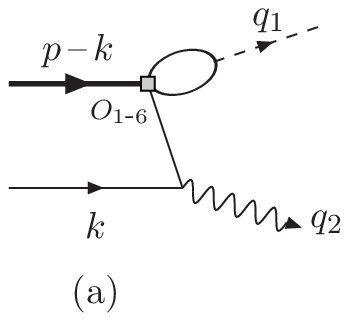}
\includegraphics[bb=80 670 190 760]{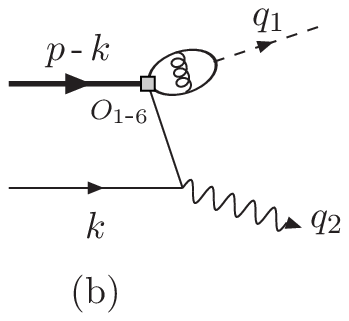}
\includegraphics[bb=80 670 190 760]{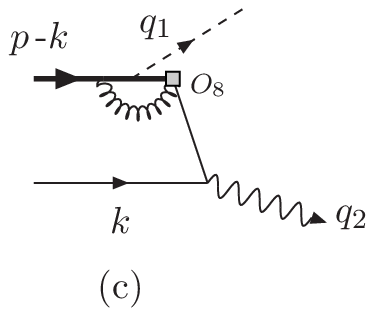}}\\
\mbox{\includegraphics[bb=80 670 190 760]{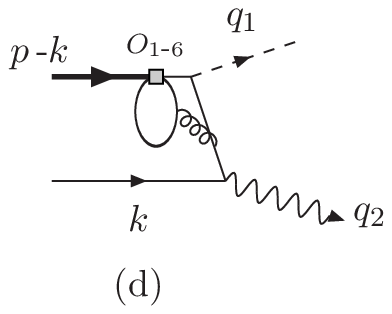}
\includegraphics[bb=80 670 190 760]{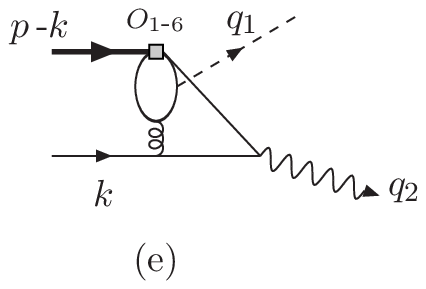}}
\end{center}
\vspace{5pt} \caption{Sample diagrams which contribute to
$\cseveneff$ ((a)\,--\,(d)). The non-factorizable diagram (e) does
not contribute at leading twist. \label{fig:aux}}
\end{figure}

We now briefly illustrate how the leading-twist contributions
proportional to $C_{1\textrm{\,--\,}6}$ and $C_{8}$ are absorbed
into $\cseveneff$. It is not surprising that the diagram in
fig~\ref{fig:aux}(a) has the structure of fig.~\ref{fig:tree},
where the grey circle represents the electromagnetic operator
$O_7$, and its contribution can therefore be included into
$\cseveneff$. Indeed this is conventionally done~\cite{effham}. In
fig.~\ref{fig:aux}(b), (c) and (d) we draw, for illustration,
three diagrams at $O(\alpha_s)$ contributing to $\cseveneff$. The
key point to note is that the loop integrations are dominated by
the short-distance region of phase space, and therefore the
contributions to $\cseveneff$ are calculable in perturbation
theory. In fact $\cseveneff$ at $O(\alpha_s)$ can be deduced from
the calculations in refs.~\cite{aagw,bfs} (for the penguin
operators $O_{3\textrm{\,--\,}6}$, for which the Wilson
coefficients are small, the contribution to $\cseveneff$ is only
known at $O(\alpha_s^0)$). In the basis and notation of
refs~\cite{aagw,bfs}
\begin{equation}
\cseveneff=C_7-\frac{C_3}{3}-\frac{4C_4}{9}-\frac{20\,C_5}{3}-\frac{80\,C_6}{9}-
\frac{\alpha_s}{4\pi}\sum_{1,2,8}C_iF_i^{(7)}\,,
\end{equation}
where $F_{1,2}^{(7)}$ and $F_8^{(7)}$ are given in eq.~(56) of
ref.~\cite{aagw} and eq.~(82) of ref.~\cite{bfs} respectively.
There are other contributions proportional to
$C_{1\textrm{\,--\,}6}$ and $C_{8}$ which cannot be absorbed into
$\cseveneff$. An example is drawn in fig.~\ref{fig:aux}(e).
However, power-counting arguments show that these contributions
are suppressed by at least one power of $M_B$. The contributions
to the hard-scattering kernel from the $O(\alpha_s)$ diagrams
drawn in fig.~\ref{fig:oneloop}, where the grey circle represents
the tree-level insertion of $O_7$, are evaluated explicitly in
this letter.

The amplitude (${\cal A}$) for the decay $\bgg$ can be written in
terms of invariant form factors $A_{\pm}$:
\begin{equation}
{\mathcal{A}}(B\to\gamma\gamma)
 =\frac{\alpha G_F f_B}{3\pi\sqrt{2}}
  \Big\{2
  A_-\,\varepsilon(q_1,q_2,\epsilon^*_1,\epsilon^*_2)+
  A_+\, \left[\,2i(q_1\cdot\epsilon^*_2)(q_2\cdot\epsilon^*_1)
           -iM_B^2(\epsilon^*_1\cdot\epsilon^*_2)\,\right]\Big\}.
\end{equation}

Following ref.~\cite{dgs2}, we choose the external state
$H=b\bar{q}$ to determine the contribution to the hard-scattering
kernel. We denote the contribution from the tree diagram in
fig.~\ref{fig:tree}, where the photon with momentum $q_1$ is
emitted from the effective vertex, to the
$\langle\gamma\gamma|{\cal H}|b\bar q\rangle$ matrix element by
$F_1^{(0)}$ (0 denotes tree-level or ``zero loops"). Similarly
$F_2^{(0)}$ is the contribution with the photon with momentum
$q_2$ emitted from the effective vertex, and is equal to
$F_1^{(0)}$ with the obvious replacements $1\leftrightarrow 2$ and
$+\leftrightarrow -$ in the expressions below.

Evaluating the diagram we find for the matrix element at tree
level
\begin{equation}
F_1^{(0)} = \lambda_t^{(q)} \frac{eG_F}{8\pi^2\sqrt{2}} m_b
\cseveneff
     \cdot \frac{e_q}{\sqrt{2}M_B k_+}
  \bar{v}^s(k)\left[\dirac{\epsilon}_2^*\dirac{q}_2
[\dirac{q}_1,\dirac\epsilon_1^*](1+\gamma_5)\right]u^S(p-k),
\end{equation}
where $u$ and $v$ are the wave functions of the $b$ and $\bar q$
quarks with spin labels $S$ and $s$. Writing this amplitude as a
convolution
\begin{equation}
F_1^{(0)}=\int \frac{d\tp}{2\pi} \Phi^{b\bar q\,(0)}_{\alpha\beta}
   T_{1\,\beta\alpha}^{(0)}
\end{equation}
(with $\Phi^{b\bar q}$ defined in eq.~(\ref{eq:phihdef}) with
$H=b\bar q$), the contribution to the hard-scattering kernel is
\begin{equation}
T_{1\,\beta\alpha}^{(0)}= \lambda_t^{(q)}
\frac{eG_F}{8\pi^2\sqrt{2}} m_b \cseveneff
     \cdot \frac{e_q}{\sqrt{2}M_B \tp}
        \left[\dirac{\epsilon}_2^*\dirac{q}_2
        [\dirac{q}_1,\dirac\epsilon_1^*](1+\gamma_5)\right]_{\beta\alpha}
\,.
\label{eq:tgg0}
\end{equation}
Adding the contribution from
$T_{2\,\beta\alpha}^{(0)}$, we obtain for the lowest-order
contribution to the two form factors:
\begin{eqnarray}
A_+&=&A_-=\int d\tp \Phi_+^B(\tp;\mu_F) T^{(0)}(\tp;\mu_R,\mu_F)
      =\lambda_t^{(q)} \cseveneff (\mu_R) \frac{M_B} {\lambda_B}\,,\\
T^{(0)}&=& \cseveneff (\mu_R) \frac{\lambda_t^{(q)}
M_B}{\sqrt{2}\tp}\,, \qquad \frac{1}{\lambda_B}=\frac{1}{\sqrt{2}}
    \int\,d\tpm\, \frac{\Phi_+^B(\tpm)}{\tpm}\,.
\label{eq:tbgg0}
\end{eqnarray}
Note that $\lambda_B$ is also the nonperturbative quantity which
is needed in evaluating the contribution from tree-level
hard-spectator interactions to exclusive nonleptonic $B$-decay
amplitudes~\cite{BBNS1,BBNS2,BBNS3}.

At one-loop order, the contribution of the diagram in
fig.~\ref{fig:oneloop}(b) to the $\langle\gamma\gamma|{\cal
H}|b\bar u\rangle$ matrix element is
\begin{equation}
F_1^{(1)\,\textrm{\scriptsize wk}}=F_1^{(0)}\frac{\as C_F}{4\pi}
   \left[-\log^2\frac{\sqrt{2} k_+}{M_B}
      -2 \log\frac{\sqrt{2} k_+}{M_B} - \pi^2\right].
\label{eq:f1wk}\end{equation} In order to determine the one-loop
contribution to $T$ ($T^{(1)}$), we need to subtract the
corresponding contribution to $\Phi^{b\bar q\,(1)}\otimes
T^{(0)}$, the convolution of the one-loop component of
$\Phi^{b\bar u}$ with $T^{(0)}$:
\begin{equation}
\left(\Phi^{(1)\,\textrm{\scriptsize wk}}\otimes T^{(0)}\right)_1
  = F_1^{(0)} \cdot \frac{\as C_F}{4\pi}
\left[-\frac{4}{\omega^2}+\frac{2}{\omega}\log\frac{2k_+^2}{\mu_F^2}
        -2\log^2\frac{\sqrt{2}k_+}{\mu_F}-\frac{3\pi^2}{4}\right]\,,
\label{eq:phi1wk}\end{equation} where we use dimensional
regularization, working in $4-\omega$ dimensions. We recall that
the subscript 1 denotes the contribution from the graphs with the
photon of momentum $q_1$ emitted from the effective vertex.

The contribution to the hard-scattering kernel from the diagram in
fig.~\ref{fig:oneloop}(b) with $q_1\leftrightarrow q_2$ is the
same, since the resulting convolution over $\tm$ is equal by
symmetry to the one over $\tp$ from fig.~\ref{fig:oneloop}(b). We
therefore present our answers below in terms of a convolution with
$\tp$ as the (dummy) integration variable, and our expressions for
the hard-scattering kernel correspond to the sum of the two
contributions.

The results in eqs.~(\ref{eq:f1wk}) and (\ref{eq:phi1wk}), together
with those in ref.~\cite{dgs2} for the remaining diagrams in
fig.~\ref{fig:oneloop} can be combined to give the one-loop
expression for the hard-scattering kernel:
\begin{eqnarray}
T&=&T^{(0)}\Bigg[1+\frac{\as C_F}{4\pi}
   \Bigg(\log^2\frac{\sqrt{2}\tp}{M}
    +4\log\frac{\sqrt{2}\tp}{M}\log\frac{M}{\mu_F}
    \nonumber\\
&&\hspace{-0.3in}
+2\log^2\frac{M}{\mu_F}+7\log\frac{M}{\mu_F}-2\log\frac{\mu_R}{\mu_F}-7
    -\frac{\pi^2}{4}\Bigg)\Bigg]\,,
\label{eq:t1gg}\end{eqnarray} where the explicit expression for
$T^{(0)}$ is given in eq.~(\ref{eq:tbgg0}).

We now use the SCET formalism to resum the large logarithms
following the steps in sec.~5 of ref.~\cite{dgs2}. The contribution
of the tensor operator to the $\langle\gamma\gamma|{\cal H}|b\bar
u\rangle$ matrix element (i.e. the diagrams in figs.~\ref{fig:tree}
and \ref{fig:oneloop}(b) plus half of the diagrams
in figs.~\ref{fig:oneloop}(d) and \ref{fig:oneloop}(e)) is
\begin{eqnarray}
&&F^{J_W}=F_1^{(0)}\Bigg\{1+ \frac{\as C_F}{4\pi}
\left[2\log\frac{M_B}{\mu_R}-\log^2\frac{\sqrt{2}k_+}{M_B}
 -\frac{7}{2}
     \log\frac{\sqrt{2}k_+}{M_B}
     -\frac{5}{2}-\pi^2 \right]\bigg\}\nonumber\\
 && +F_2^{(0)}\bigg\{1+ \frac{\as C_F}{4\pi}
\left[2\log\frac{M_B}{\mu_R}-\log^2\frac{\sqrt{2}k_-}{M_B}
 -\frac{7}{2}
     \log\frac{\sqrt{2}k_-}{M_B}
     -\frac{5}{2}-\pi^2 \right]\Bigg\}\,.\label{eq:fjw}
\end{eqnarray}
In the SCET formalism, only one operator, denoted by $O_9$ in
ref.~\cite{SCETbtou}, contributes to the form factors. From
eq.~(66) of ref.~\cite{dgs2} and the neighbouring discussion, we
can rewrite $F^{J_W}$ in terms of the SCET Wilson coefficient
$C_9^{SCET}(E_\gamma;\mu_R,\mu_F)$ as
\begin{eqnarray}
F^{J_W}&=& C^{SCET}_9\!\big(M_B/2;\mu_R,M_B\big)\times\nonumber
\\
&& \Bigg\{F_1^{(0)}\bigg(1+ \frac{\as C_F}{4\pi}
 \big[
     -\log^2\frac{\sqrt{2}k_+}{M_B}
     -\frac{7}{2}\log\frac{\sqrt{2}k_+}{M_B}
     +\frac{7}{2}-\frac{11\pi^2}{12}\big]\bigg)\nonumber\\
&&\hspace{-0.5in}+F_2^{(0)}\bigg(1+ \frac{\as C_F}{4\pi}
 \big[
     -\log^2\frac{\sqrt{2}k_-}{M_B}
     -\frac{7}{2}\log\frac{\sqrt{2}k_-}{M_B}
     +\frac{7}{2}-\frac{11\pi^2}{12}\big]\bigg)\Bigg\}\,.
\label{eq:fjwscet}\end{eqnarray} Comparing eqs.~(\ref{eq:fjw}) and
(\ref{eq:fjwscet}) we find
\begin{equation}
C_9^{SCET}(\frac{M_B}{2};\mu_R,M_B)
   = 1+\frac{\as C_F}{4\pi}\left[
   2\log\frac{M_B}{\mu_R}-6-\frac{\pi^2}{12}
   \right]\,,
\end{equation}
which agrees with eq.~(33) of ref.~\cite{SCETbtou} for $\mu_R=M_B$.

Finally, we can use the renormalization group equation for
$C_9^{SCET}$, derived in ref.~\cite{SCETbtou}, to obtain its value
at any factorization scale $\mu_F$. We conclude that the form
factors $A_+=A_-$ are given by the generic form in
eq.~(\ref{eq:fgeneric}) with
\begin{equation}
T(\tp;\mu_R,\mu_F)=
\frac{\lambda_t^{(q)}M_B}{\sqrt{2}\tp}\,\cseveneff (\mu_R)\,
   C^{SCET}_9(M_B/2\hspace{2pt};\mu_R,\mu_F)\,
\left[1+\frac{\as C_F}{4\pi} K_t\big(\tp,M_B/2;\mu_F\big)
   \right].   \label{eq:tbgg}
\end{equation}
Thus the hard-scattering kernel has exactly the same dependence on
$\tp$ as that for $\bwg$ decays in eq.~(\ref{eq:previous}).

\subsection{The decay \boldmath{$\bllg$}}

The procedure for the evaluation of the $\bllg$ decay amplitude
follows exactly the same steps. We will assume here that
$x_\gamma\equiv 2E_\gamma/M_B$ satisfies
$(1-x_\gamma)\gg\lqcd/M_B$. This allows us to neglect the diagrams
in which the real photon is emitted from the weak vertex and the
virtual one is radiated from the light quark. The effective
Hamiltonian for this process is then~\cite{effham,bgllold,bgll}:
\begin{equation}
{\mathcal{H}}= \lambda_t^{(q)} \frac{\alpha G_F}{\pi\sqrt{2}}
 \bigg(
  C_9^{\rm eff} (\bar{q}\gamma_\mu b_L)(\bar\ell \gamma^\mu \ell)
+C_{10} (\bar{q}\gamma_\mu b_L)(\bar\ell\gamma^\mu\gamma_5 \ell)
-2 \frac{\cseveneff}{q_1^2}m_b
   (\bar{q}i\sigma_{\mu\nu} q_1^\nu b_R)(\bar\ell\gamma^\mu \ell)
 \bigg)\,,\label{eq:hbllg}
\end{equation}
where $b_{L,R}=(1\mp \gamma_5)/2\,b$. Again $C_9^{\rm eff}$
contains short-distance, leading-twist contributions from
four-quark and chromomagnetic operators:
\begin{equation}
C_9^{\rm eff}=C_9 + Y(q_1^2)
-\frac{\alpha_s}{4\pi}\sum_{1,2,8}C_iF_i^{(9)}\,,
\end{equation}
where $Y(q^2_1)$ has contributions from $O_{1\textrm{\,--\,}6}$
and its explicit expression is given in eq.~(10) of
ref.~\cite{bfs} and $F_{1,2}^{(9)}$ and $F_{8}^{(9)}$ are given in
eqs.~(54) and (55) of ref.~\cite{aagw} and eq.~(83) of
ref.~\cite{bfs} respectively. We follow the authors of
ref.~\cite{bgll} in neglecting the long-distance effects
associated with $c\bar c$ resonances. For the purposes of this
paper, we start with eq.~(\ref{eq:hbllg}) and evaluate the matrix
elements of the operators on the right-hand side. The four form
factors for this process are defined by:
\begin{equation}
2\langle\gamma(q,\epsilon^*)| \bar{q}\gamma_\mu
b_L|\bar{B}(p)\rangle = e \epsilon_{\mu\nu\rho\sigma}
{\epsilon^*}^\nu v^\rho q^\sigma F_V(\eg)+
    ie[\epsilon^*_\mu (v\cdot q) -q_\mu (v\cdot \epsilon^*)] F_A(\eg)
\end{equation}
and
\begin{eqnarray}
2\langle\gamma(q,\epsilon^*)| \bar{q}i\sigma_{\mu\nu}(p-q)^\nu
       b_R|\bar{B}(p)\rangle &=& -eM_B \epsilon_{\mu\nu\rho\sigma} {\epsilon^*}^\nu v^\rho
q^\sigma F_T(\eg)\nonumber\\
&&\hspace{-0.8in}    - ieM_B[\epsilon^*_\mu (v\cdot q) -q_\mu
(v\cdot \epsilon^*)]
    F'_T(\eg)\,.
\end{eqnarray}
Since the corrections to the vector and axial currents have
already been discussed in ref.~\cite{dgs2}, we consider here only
the form factors $F_T$ and $F_T^\prime$. Evaluating the diagram in
fig.~\ref{fig:tree} we find that at lowest order the contribution
to the matrix element $2\langle \ell^+\ell^-\gamma|\bar
qi\sigma_{\mu\nu}q_1^\nu b_R|b\bar q\rangle$ is
\begin{equation}
F^{(0)}_\mu =\frac{e_q}{2(q_2)_- k_+}\,
     \bar{v}^s(k)
        \left[\dirac{\epsilon}^*\dirac{q}_2
            [\gamma_\mu,\dirac{q}_1]\right]u^S_R(p-k)\,,
\label{eq:fmu0}\end{equation} and the hard-scattering amplitude is
\begin{equation}
T^{(0)}_\mu=\frac{e_q}{4(q_2)_- k_+} \left[
     \dirac{\epsilon}^*\dirac{q}_2
            [\gamma_\mu,\dirac{q}_1](1+\gamma_5)
  \right]_{\beta\alpha}\,.
\end{equation}
The form factors at lowest order are thus given by
\begin{equation}
F_T = F'_T = \frac{Q_q f_B M_B}{2\eg \lambda_B}\,,
\end{equation}
where $Q_q=1/3$ is the electric charge of $\bar{q}$.

At one-loop order the contribution
from the diagram of fig.~\ref{fig:oneloop}(b)
to the matrix element $2\langle
\ell^+\ell^-\gamma|\bar qi\sigma_{\mu\nu}q_1^\nu b_R|b\bar
q\rangle$ is
\begin{equation}
F^{(1)\,\textrm{\scriptsize wk}}_\mu=F_\mu^{(0)}\cdot\frac{\as
C_F}{4\pi}
  \Bigg[-\log^2\frac{z}{y}-2\log\frac{z}{y}-
2\log\frac{z}{y}\log\frac{x}{y}     -\log^2\frac{x}{y}
      -2{\rm Li}_2\left(1-\frac{y}{x}\right)
      -\pi^2\Bigg]\,,
\end{equation}
where $x=M_B^2$, $y=2M_BE_\gamma$ and $z=2q_2\cdot k$. The
corresponding contribution to $\Phi^{b\bar q\,(1)}\otimes T^{(0)}$
is given by the r.h.s. of eq.~(\ref{eq:phi1wk}) with $F_1^{(0)}$
replaced by $F_\mu^{(0)}$ given in (\ref{eq:fmu0}).

Combining the above results with the remaining diagrams in
fig.~\ref{fig:oneloop} from ref.~\cite{dgs2} we find that
the hard-scattering kernel is given by
\begin{eqnarray}
T_\mu&=&T_\mu^{(0)}\Bigg[1+\frac{\as C_F}{4\pi}\Big(
\log^2\frac{z}{x}+2\log\frac{z}{x}\log\frac{x}{\mu_F^2}
-2\log\frac{\mu_F^2}{y}
-2\log^2\frac{x}{y}\nonumber\\
&&\qquad +\frac{1}{2}\log^2\frac{x}{\mu_F^2} -2{\rm
Li}_2\left(1-\frac{y}{x}\right)
-\log\frac{\mu_R^2}{\mu_F^2}+\frac{3}{2}\log\frac{x}{\mu_F^2}
-\frac{\pi^2}{4}-7 \Big)\Bigg]\,.
\end{eqnarray}

We now perform the resummation of the large logarithms as for the
$\bgg$ decays above. By comparing our results for the contribution
of the tensor operator to the matrix element $F^{J_W}$ with the
results for the same transition in the SCET, we obtain the
expression of $C_9^{SCET}(\eg;\mu_R,\mu_F=M_B)$ (it is given
explicitly in eq.~(\ref{eq:c9scet}) below). For $\mu_R=M_B$ our
result agrees with eq.~(33) of ref.~\cite{SCETbtou}.

In this way we find that the form factors $F_T$ and $F_T^\prime$
are given by the generic form in eq.~(\ref{eq:fgeneric}) with
\begin{equation}
T(\tp,\eg;\mu_R,\mu_F)=
\frac{f_B M_B}{6\sqrt{2}\eg}\frac{1}{\tp}
   C^{SCET}_9(\eg;\mu_R,\mu_F)
\left[1+\frac{\as C_F}{4\pi} K_t(\tp,\eg;\mu_F)\right]\,.
\end{equation}

This completes the demonstration that at one-loop and
leading-twist order the hard-scattering kernel has the same form
for all three radiative decays.

\section{Discussion and Conclusions\label{concs}}

As a consequence of the universality of the nonperturbative
effects in radiative $B$-decays the rates can be related simply
using perturbation theory. The decay rates are given by the
following expressions:
\begin{eqnarray}
\Gamma(\bar{B}_q\to\gamma\gamma)&=&\frac{\alpha^2 G_F^2M_B^5
f_B^2}{144\pi^3}\, \left(C_7^{{\rm eff}}\right)^2\,
|\lambda_t^{(q)}|^2
  (C^{SCET}_9)^2\,\frac{1}{\Lambda_B^2(M_B/2)}\,,\label{eq:ratebgg}\\
\frac{d\Gamma(\bar{B}_q\to\gamma e^+ e^-)}{d\eg}
 &=&\frac{\alpha^3 G_F^2 M_B^4 f_B^2}{1728\pi^4} |\lambda^{(q)}_t|^2
     \,\frac{x_\gamma(1-x_\gamma)}{\Lambda_B^2(\eg)}\times
\nonumber\\
&& \left[\Big|\,C_9^{\rm eff}\,C^{SCET}_3
        +\frac{2\,\cseveneff}{1-x_\gamma}\,
         C^{SCET}_9\,\Big|^2
      +\left|\,C_{10}\,C^{SCET}_3\,\right|^2\right]\!,\label{eq:ratebllg}\\
\frac{d\Gamma(B^+\to\gamma e^+\nu)}{d\eg}
  &=&\frac{\alpha G_F^2 f_B^2 |V_{ub}|^2 M_B^4}{54\pi^2} (C^{SCET}_3)^2
  \frac{x_\gamma(1-x_\gamma)}{\Lambda_B^2(\eg)}\,,
\label{eq:ratebwg}\end{eqnarray} where $x_\gamma=2E_\gamma/M_B$.
$C_{3,9}^{SCET}$ in eqs.~(\ref{eq:ratebgg})\,-\,(\ref{eq:ratebwg})
denote the matching coefficients relating the QCD and SCET
operators evaluated at $\mu_F=M_B$ and are calculable in
perturbation theory,
\begin{equation}\label{eq:c9scet}
C^{SCET}_9
   = 1+\frac{\as(M_B) C_F}{4\pi}
\bigg(\log\frac{M_B^2}{\mu_R^2}
   -2\log^2 x_\gamma+2\log x_\gamma
   -2{\rm Li}_2\left(1-x_\gamma\right)
   -6-\frac{\pi^2}{12}
   \bigg)
\end{equation}and
\begin{equation}
C^{SCET}_3
   = 1+\frac{\as(M_B) C_F}{4\pi}\bigg(
   -2\log^2 x_\gamma
   -2{\rm Li}_2\left(1-x_\gamma\right)
   +\frac{3x_\gamma-2}{x_\gamma-1}\log x_\gamma
   -6-\frac{\pi^2}{12}
   \bigg)\,.
\end{equation}
$\mu_R$ is the scale where the Wilson coefficient functions of the
operators in the effective Hamiltonian, $\cseveneff,\,C_9^{\rm
eff}$ and $C_{10}$, are evaluated, and for convenience we set
$\mu_R=M_B$. The only nonperturbative parameter in
eqs.~(\ref{eq:ratebgg})\,-\,(\ref{eq:ratebwg}) is
$\Lambda_B(E_\gamma)$, which is the generalization of $\lambda_B$:
\begin{equation}\label{eq:LambdaB}
\frac{1}{\Lambda_B(\eg)}=\frac{1}{\sqrt{2}}\exp[-S(E_\gamma;\mu_F)]
   \int d\tp \frac{\Phi_+^B(\tp;\mu_F)}{\tp}\left[1+\frac{\as C_F}{4\pi}
  \left(\log^2\frac{2\sqrt{2}\eg\tp}{\mu_F^2}
   -\frac{\pi^2}{6}-1\right)\right]\,.
\end{equation}
The resummation of large (Sudakov) logarithms is obtained through
the evolution of SCET operators from $M_B$ down to $\mu_F$ (this
evolution is common to all the operators), leading to the
exponential factor in $\Lambda_B(\eg)$
\begin{eqnarray}
-S(E_\gamma;\mu_F)&=& \frac{f_0(r)}{\as(m_b)}+f_1(r)\,,\\
f_0(r)&=&-\frac{4\pi C_F}{\beta^2_0}\left[\frac{1}{r}-1+\log
r\right]\,,\\
f_1(r)&=&-\frac{C_F\beta_1}{\beta_0^3}\left[1-r+r\log
r-\frac{\log^2 r}{2}\right]\nonumber\\ &&
+\frac{C_F}{\beta_0}\left[\frac{5}{2}-2\log x_\gamma\right]\log r
-\frac{2C_F B}{\beta_0^2}[r-1-\log r]\,,
\end{eqnarray}
where $r=\as(\mu_F)/\as(m_b)$, $\beta_0=11C_A/3 -2N_f/3$ and
$\beta_1=34C_A^2/3-10C_A N_f/3 -2C_F N_f$.

\begin{figure}[t]
\begin{center}
\includegraphics[angle=270,width=9cm]{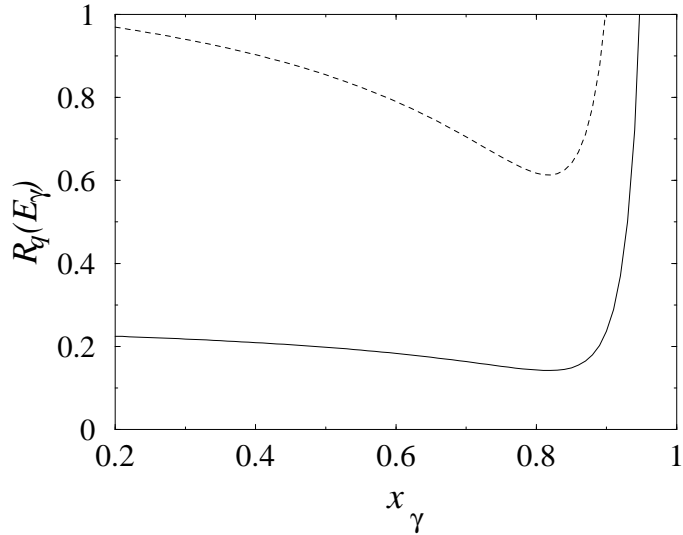}
\caption{Ratios of decay rates $R_{d}$ (solid line $\times
10^{-4}$) and $R_s$ (dashed line $\times 10^{-3}$) as functions of
$x_\gamma=2\eg/M_B$. \label{fig:ratio}}
\end{center}
\end{figure}

In this letter we do not perform a detailed phenomenological
analysis. For illustration however, in fig.~\ref{fig:ratio} we
plot the ratio
\begin{equation}
R_q(\eg)=\frac{d\Gamma(\bar{B}_q\to\gamma e^+ e^-)/d\eg}
   {d\Gamma(B^+\to\gamma e^+ \nu)/d\eg}\,,
\end{equation}
for $q=d,s$ and for the following values for the Wilson
coefficients: $\cseveneff=-0.390$, $C_9^{\rm eff}=4.182$ and
$C_{10}=-4.234$~\cite{effham,bgllold,bgll} (a more detailed
analysis would include the effects of $Y(q_1^2)$ and the
$O(\alpha_s)$ corrections to these coefficients, but these are
sufficiently small to be neglected in this presentation). We take
$|\lambda_t^{(d)}|=7.8\times 10^{-3}$,
$|\lambda_t^{(s)}|=4.1\times 10^{-2}$, $|V_{ub}|=3.6\times
10^{-3}$, $f_{B_d}=190\,$MeV and $f_{B_s}=230\,$MeV. In other
words we plot
\begin{eqnarray}
R_d\times\left(\frac{7.8\times
10^{-3}}{|\lambda_t^{(d)}|}\right)^2
      \!\! \left(\frac{|V_{ub}|} {3.6\times 10^{-3}}\right)^2
\!\!\left(\frac{f_{B_d}}{190\,\textrm{MeV}}\right)^2
\qquad\textrm{and}\\
R_s\times\left(\frac{4.1\times
10^{-2}}{|\lambda_t^{(s)}|}\right)^2 \!\! \left(\frac{|V_{ub}|}
{3.6\times
10^{-3}}\right)^2\!\!\left(\frac{f_{B_s}}{230\,\textrm{MeV}}\right)^2\!\!.
\end{eqnarray}
The two curves are proportional to each other, differing by the
values of the leptonic decay constants and CKM matrix elements.
They do not depend however, on the model for the distribution
amplitude. As $x_\gamma$ approaches 1 the virtual photon
approaches its mass shell and this is the reason for the rise of
the curves in this region. We recall however that our expressions
are valid in the region $1-x_\gamma\gg\lqcd/M_B$ (and
$x_\gamma\gg\lqcd/M_B$).

Knowledge of $\Lambda_B$, which we do not have at present, is
required to determine each of the branching ratios separately.
Taking $\Lambda_B(M_B/2)=350$\,MeV, we estimate that the $\bgg$
branching ratios are approximately
\begin{equation}
B(B_d\to\gamma\gamma)\simeq 3\times
10^{-8}\qquad\textrm{and}\qquad B(B_s\to\gamma\gamma)\simeq
10^{-6}\,,
\end{equation}
of the same order of magnitude as in ref.~\cite{bgg}.  For the
remaining decays we estimate:
\begin{eqnarray}
B(B^+\to\gamma\ell^+\nu)&=&O(10^{-6})\,,\\
B(B_s\to\gamma\ell^+\ell^-;\, x_\gamma<0.8)&=&O(10^{-9})\,,\\
B(B_d\to\gamma\ell^+\ell^-;\,
x_\gamma<0.8)&=&O(10^{-11})\,\textrm{--}\ O(10^{-10})\,,
\end{eqnarray}
where we have imposed the cut-off $x_\gamma<0.8$ in order to
satisfy the constraint $(1-x_\gamma)\gg\lqcd/M_B$. Assuming that
there is no very large enhancement as $x_\gamma\to 1$, the result
for $B(B_s\to\gamma\ell^+\ell^-)$ is in broad agreement with the
estimates in ref.~\cite{bgllold}. This result is an order of
magnitude smaller than that of ref.~\cite{bgll}. The discrepancy
is partly due to the fact that we do not reproduce the formula for
the ratio $B(B_s\to\gamma\ell^+\ell^-)/B(B_s\to\gamma\gamma)$ used
in ref.~\cite{bgll}, partly to a different choice of input
parameters, partly due to the fact that the authors of \cite{bgll}
integrate up to the kinematic limit for $x_\gamma$ and partly to
the fact that we include higher-order QCD corrections.

The explicit calculations described in the previous section, were
performed at one-loop order. Nevertheless, as we now explain, we
anticipate that the general features will survive also at higher
orders. For all the decays the lowest-order diagram of
fig.~\ref{fig:tree} gives a factor of $1/k_+$ which leads to the
factor of $1/\tp$ in the hard-scattering kernels. At one-loop
order all the diagrams in fig.~\ref{fig:oneloop} except for
\ref{fig:oneloop}(c) have a collinear light-quark propagator
external to the loop giving a factor of $1/k_+$. In order for the
diagram in fig.~\ref{fig:oneloop}(c) to contribute at leading
twist we must recover the factor of $1/k_+$ from the loop
integration. As explained in ref.~\cite{dgs2}, such a contribution
comes from the soft region of phase space and is absorbed into the
wave function -- it cancels when $\Phi^{(1)}\otimes T^{(0)}$ is
subtracted, and therefore does not contribute to the
hard-scattering kernel. Thus the one-loop contribution to the
hard-scattering kernel has two components:
\begin{itemize}
\item the correction to the electromagnetic vertex --
figs.~\ref{fig:oneloop}(a),(e) and half of (d)),
\item the correction to the weak vertex --
figs.~\ref{fig:oneloop}(b),(f) and half of (d)).
\end{itemize}
The power-counting arguments leading to this structure appear to
be sufficiently general for us to conjecture that it is true to
all orders. The corrections to the electromagnetic vertex are then
clearly common to all three decay processes. Even though the
corrections to the weak vertex are not common, the differences
only arise at scales of $O(M_B)$. For soft and collinear gluons,
the Dirac structure of the heavy and light quarks is indeed such
that the result is independent of the form of the weak operator,
and of the $\gamma$-matrix (matrices) in particular. Although
there are two large scales in each of these decay processes, $M_B$
and $\sqrt{M_B\lqcd}$, the differences are only due to physics at
the scale $M_B$ and can be expressed in terms of a calculable
perturbation series in $\alpha_s(M_B)$. Recent studies of
factorization at higher orders of perturbation theory include a
demonstration of the validity of eq.(\ref{eq:fgeneric}) for $\bwg$
decays~\cite{lpw} and a study of the transitions from soft to
collinear quarks in the SCET~\cite{hillneubert}.

The processes studied in this letter have been rare radiative
decays. For other processes, such as the two-body exclusive decays
of B-mesons into two light mesons~\cite{BBNS1,BBNS3}, the
hard-scattering kernels will in general be different.
Nevertheless, as a result of the independence of soft and
collinear QCD effects from the structure of the weak operator, it
is to be expected that there are analogous relations between
contributions to different decay amplitudes, and we are currently
investigating the structure and scope of these relations.

We end by repeating that in this letter we have only considered
the leading-twist contributions to radiative $B$-decays. For a
detailed and precise phenomenology of exclusive $B$-decays the
extension of the factorization formalism to the $O(\lqcd/M_B)$
corrections will be necessary (see, for example,
refs.~\cite{Benekepower,Pirjolpower} for recent contributions
towards this).

\section*{Acknowledgements}
We thank M.~Beneke, M.~Neubert and L.~Sehgal for helpful comments
on the manuscript.

This work has been partially supported by PPARC, through grants
PPA/G/O\-/1998/00525 and PPA/G/S/1998/00530.

\end{document}